\documentclass[letterpaper,twocolumn,showpacs,prl,aps,floatfix]{revtex4}
\usepackage[latin1]{inputenc}
\usepackage{bm}
\usepackage{multirow,amssymb,amsbsy,amsmath}
\usepackage{graphicx}
\makeatletter
\usepackage{pifont}
\makeatother
\usepackage{color}
\definecolor{gray}{rgb}{0.6,0.6,0.6}
\definecolor{Green}{rgb}{.0,.8,.5}
\begin{document}

\title{Entanglement dynamics of a strongly driven trapped atom}

\author{Maryam Roghani}

\author{Hanspeter Helm}

\author{Heinz-Peter Breuer}

\affiliation{Physikalisches Institut, Universit\"at Freiburg,
Hermann-Herder-Strasse 3, D-79104 Freiburg, Germany}

\date{\today}

\begin{abstract}
We study the entanglement between the internal electronic and the
external vibrational degrees of freedom of a trapped atom which is
driven by two lasers into electromagnetically-induced
transparency. It is shown that basic features of the intricate
entanglement dynamics can be traced to Landau-Zener splittings
(avoided crossings) in the spectrum of the atom-laser field
Hamiltonian. We further construct an effective Hamiltonian that
describes the behavior of entanglement under dissipation induced
by spontaneous emission processes. The proposed approach is
applicable to a broad range of scenarios for the control of
entanglement between electronic and translational degrees of
freedom of trapped atoms through suitable laser fields.
\end{abstract}

\pacs{03.67.Bg, 03.65.Yz, 37.10.De}

\maketitle

A trapped atom driven by two laser beams under conditions of
elec\-tro\-mag\-netically-induced transparency (EIT) represents a
realistic model for current experiments on ion or atom trapping
\cite{roos,schmidt-kaler}. Quantum correlations between the
translational and the electronic degree of freedom of the trapped
atom play a crucial role  in the cooling dynamics of trapped atoms
and ions. Indeed, such correlations are responsible for the
transfer of vibrational energy of the trapped atom into excess
energy of the scattered radiation field \cite{RBH}.

Here, we explore the dynamical behavior of entanglement \cite{ABU,RHB}
 between the external vibrational degree of freedom of the trapped
atom and its internal electronic degree of freedom during the
cooling process. Solving numerically the master equation \cite{RH}
which describes the dynamics of the composite state $\rho(t)$ of
the trapped atom-laser field system, we investigate the evolution
of quantum correlations by use of the negativity \cite{VIDAL}
\begin{equation} \label{N-RHO}
 {\mathcal{N}}(t) = {\mathcal{N}}(\rho(t)) = \frac{1}{2}\Big(||\rho^{\rm T}(t)||-1\Big),
\end{equation}
where $\rho^{\rm T}(t)$ represents the partial transpose of
$\rho(t)$, and $||\cdot||$ denotes the trace norm. The negativity
${\mathcal{N}}(\rho)$ is a nonnegative function which quantifies
the degree of entanglement in a mixed quantum state $\rho$.
It vanishes for separable, classically correlated
states, and takes on a maximum value if $\rho$ is a pure,
maximally entangled state.  

In the trapped atom system the Lamb-Dicke parameter
$\eta=k/\sqrt{{2m\omega}}$ ($\hbar$=1) controls the probability of
changing the vibrational quantum number $n$ in electronic
transitions \cite{You}. For small $\eta$ the photon-recoil energy
$k^2/2m$ is an inconsiderable fraction of the trap vibrational
frequency $\omega$. In the Lamb-Dicke regime, $\eta\sqrt{n+1}\ll
1$, only the first sideband transitions, $\Delta n=\pm 1$, are
significant, their probability being proportional to
$(n\!+\!1)\eta^2$, where $n$ is the vibrational quantum number.
Even outside the Lamb-Dicke regime a rapid removal of vibrational
energy of a trapped atom occurs under suitable experimental
parameters, i.e.,  a difference of ac-Stark shifts of the two
ground states of the order of $\omega$ \cite{Morigi, RH,RBH}.
Under these conditions, with blue detuning of two
counter-propagating lasers, a vibrationally excited atom can be
driven into states with mean values $\langle n \rangle$ near zero.

\begin{figure}[h]
\includegraphics[width=0.9\columnwidth]{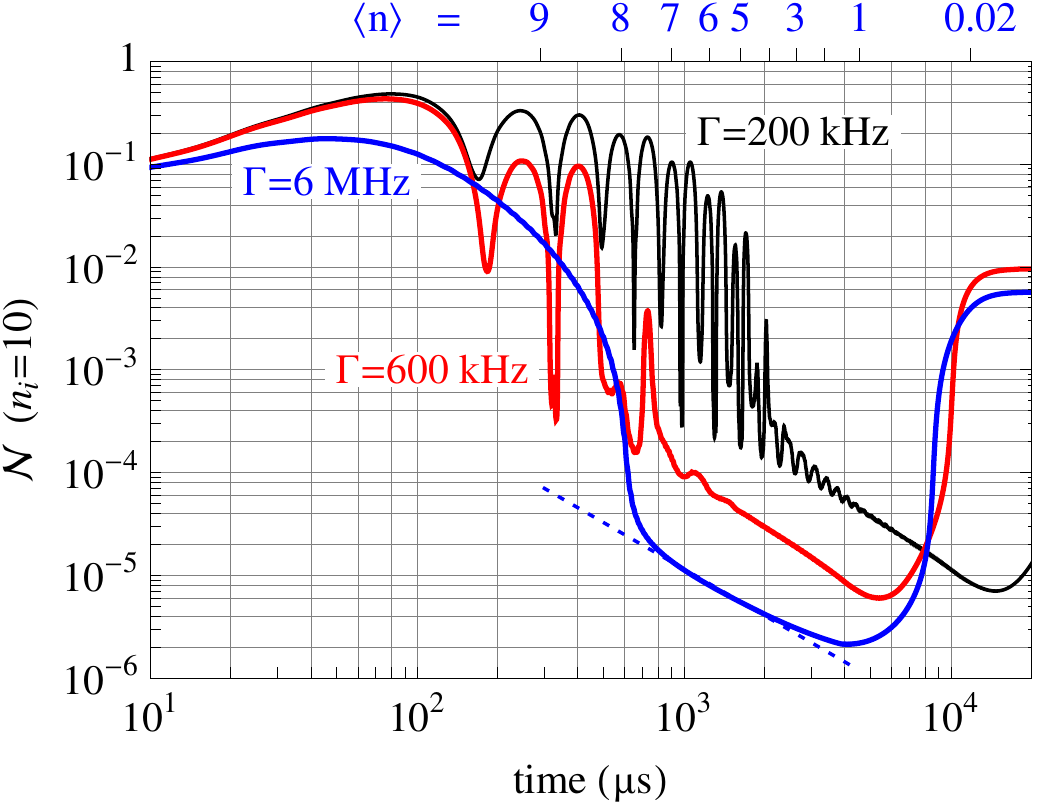}
\caption{Dynamics of the negativity during the cooling process for
the initial vibrational level $n_i=10$ for three values of the
excited state width $\Gamma$ and $\eta=0.1$. The top scale gives
the mean vibrational quantum number for the 6 MHz result. The
dotted line represents a power law decay, ${\mathcal{N}}\propto
t^{-3/2}$.} \label{figure1}
\end{figure}

An intricate behavior of the negativity is observed during such a
vibrational cooling process, as is shown in Fig.~\ref{figure1}. We
see a rapid build-up of entanglement, a subsequent exponentially
damped oscillation, followed by a precipitous drop and a power law
behavior and, finally, a rebirth of entanglement towards the
non-equilibrium stationary state.

We demonstrate that this complex behavior of the entanglement
dynamics can be related to Landau-Zener splittings or avoided
crossings in the spectrum of the atom-laser field Hamiltonian. The
dominant source of entanglement is thus a near degeneracy of
vibronic states, which appears as a consequence of the ac-Stark
shift, suitably chosen for EIT-cooling. Two-photon transitions
associated with a change in vibrational quantum number occur and
lead to a periodic oscillation of entanglement. In addition, we
derive an effective Hamiltonian that allows the modeling of the
negativity decay in the presence of 
dissipation induced by spontaneous emission processes.

\begin{figure}[h]
\includegraphics[width=0.4\columnwidth]{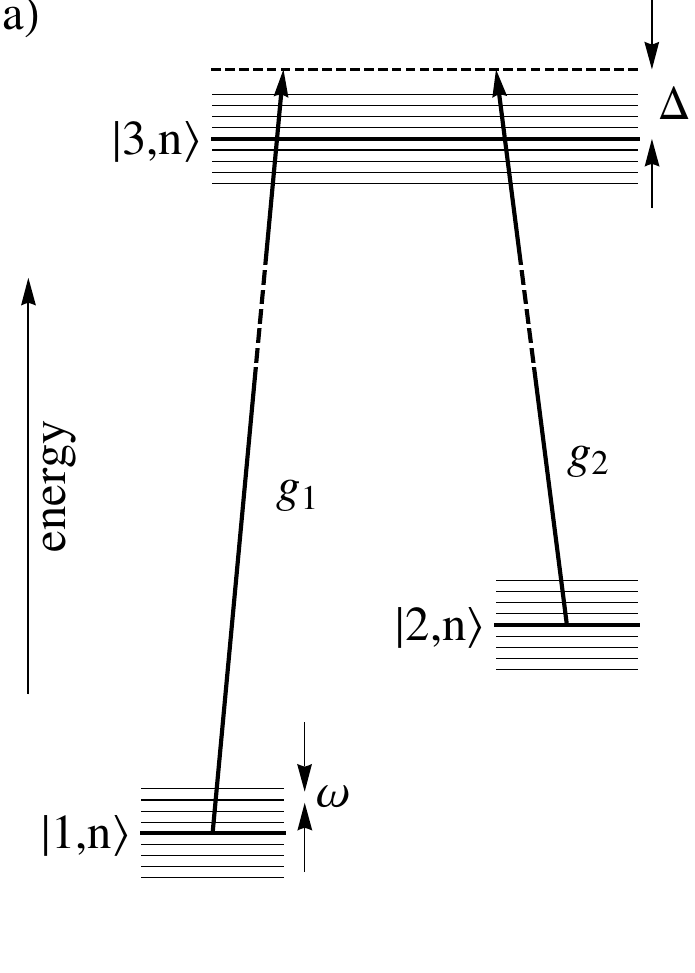}
\hspace{2mm}
\includegraphics[width=0.53\columnwidth]{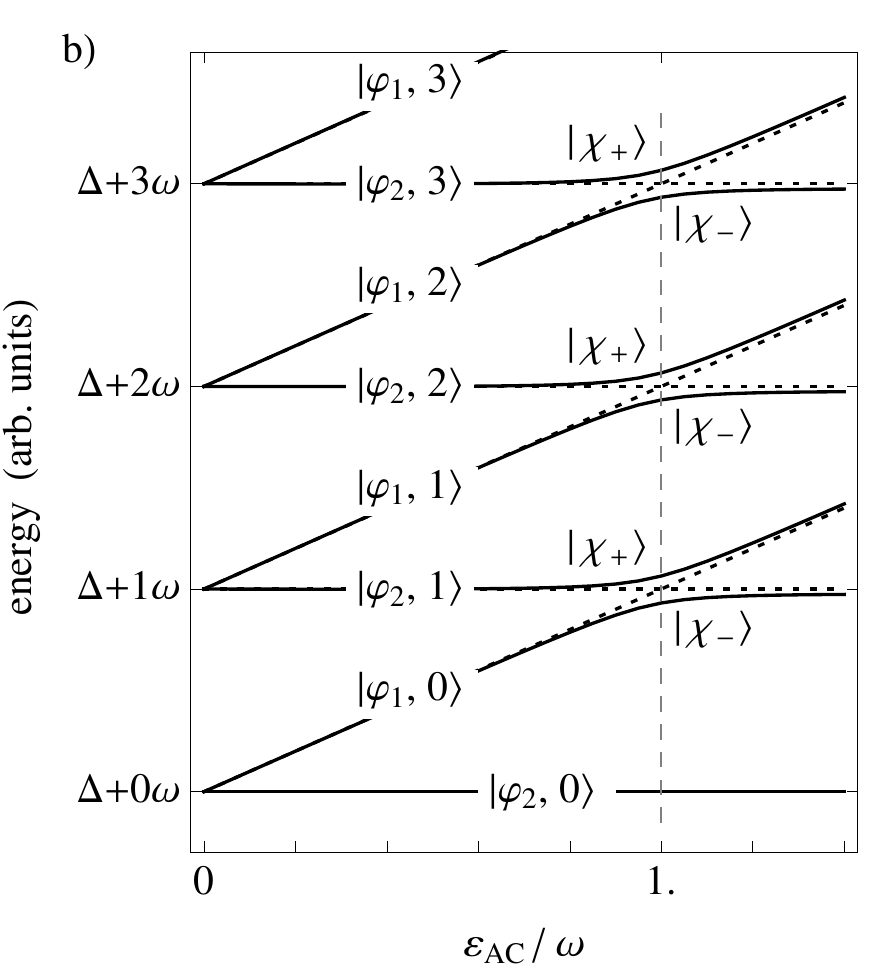}
\caption{a) Trapped 3-level atom, b) energies of dressed vibronic
states as a function of the normalized ac-Stark shift.}
\label{figure2}
\end{figure}

Our model is based on the quantum master equation
\begin{equation} \label{QMEQ}
 \frac{d}{dt}\rho(t) = -{\rm{i}}[H^{\eta},\rho(t)] + {\mathcal{L}}\rho(t)
\end{equation}
for the interaction picture density matrix $\rho(t)$ representing
the combined state of vibrational and electronic degrees of
freedom of the trapped atom. The total Hamiltonian
\begin{equation} \label{H-ETA}
 H^{\eta} = H_{\rm cm} + H_{\rm el} + H^{\eta}_{\rm int}
\end{equation}
of the model consists of three parts: $H_{\rm cm}=\omega
a^{\dagger}a$ describes the vibrational degree of freedom of a
harmonically trapped atom with raising and lowering operators
$a^{\dagger}$ and $a$, and $H_{\rm el}=\Delta (|1\rangle\langle 1|
+ |2\rangle\langle 2|)$ represents the electronic degree of
freedom with detuning $\Delta$. The electronic states are denoted
by $|i\rangle$, $i=1,2,3$, and form a $\Lambda$-type level
structure sketched in Fig.~\ref{figure2}a.
The Hamiltonian
\begin{equation}
 H^{\eta}_{\rm int} = \frac{g_1}{2} e^{ikx}|3\rangle\langle 1|
 + \frac{g_2}{2} e^{-ikx}|3\rangle\langle 2| + \rm{h.c.}
\end{equation}
models the interaction with two counter-propagating laser beams,
the most efficient configuration for cooling, where $x$ denotes
the atomic position and $k$ the wavenumber ($k_1 \approx k_2 =
k$). Note that $H^{\eta}_{\rm int}$ depends on the Lamb-Dicke
parameter $\eta$ through $kx=\eta(a^{\dagger}+a)$. Finally,
dissipation caused by spontaneous emissions is modelled by the
Lindblad superoperator (see, e.g., Ref.~\cite{BP})
\begin{equation} \label{LINDBLAD}
 {\mathcal{L}}\rho = \sum_{j=1,2} \sum_{q=\pm} \frac{\Gamma}{4}
 \left[ \sigma_{jq}\rho \sigma^{\dagger}_{jq} - \frac{1}{2}
 \left\{ \sigma_{jq}^{\dagger}\sigma_{jq}, \rho \right\} \right],
\end{equation}
where $\Gamma$ denotes the spontaneous emission rate and
$\sigma_{j+} = |j\rangle\langle 3| e^{ikx}$, $\sigma_{j-} =
|j\rangle\langle 3| e^{-ikx}$. The Lindblad operators
$\sigma_{j+}$ describe the transitions from the excited electronic
state $|3\rangle$ to the ground states $|j=1,2\rangle$ with a
photon emitted parallel to the harmonic oscillator axis, while the
$\sigma_{j-}$ describe the transitions from $|3\rangle$ to
$|j=1,2\rangle$ with a photon emitted counter-propagating to the
harmonic oscillator axis. 
This is a good compromise
short of doing a full 3D simulation as the recoil of the
translational motion is zero for emissions perpendicular to the
trap axis, while it leads to Doppler heating or cooling for
emissions along the trap axis. 
Typical results obtained by
numerically solving Eq.~\eqref{QMEQ} \cite{RH,RBH} are given in
Fig.~\ref{figure1}.

We start our discussion of the entanglement dynamics by
considering first the case $\Gamma\!=\!0$, neglecting completely
effects from spontaneous emission processes. The dynamics is then
given by the Schr\"odinger equation
\begin{equation} \label{SCHROEDINGER}
 \frac{d}{dt}|\psi(t)\rangle = -{\rm{i}}H^{\eta}|\psi(t)\rangle \, .
\end{equation}
Our goal is to describe the basic features of the dynamical
behavior of the negativity in terms of the properties of the
spectrum of the full Hamiltonian $H^{\eta}$ given by
Eq.~\eqref{H-ETA}. For vanishing Lamb-Dicke parameter, $\eta=0$,
the external and the internal degrees of freedom of the atom
decouple, such that the eigenstates of $H^0=H_{\rm cm}+H_{\rm
el}+H_{\rm int}^0$ are given by tensor product states
$|\varphi_i,n\rangle\equiv|\varphi_i\rangle\otimes|n\rangle$ with
corresponding energy eigenvalues, $\varepsilon_{i,n} =
\varepsilon_i + n \omega$. Here, $n=0,1,2,\ldots$ denotes the
vibrational quantum number. The $|\varphi_i\rangle$ are dressed
electronic states which are defined as normalized eigenstates of
$H_{\rm el}+H_{\rm int}^0$ with the eigenvalues
\begin{equation}
 \varepsilon_{1} ={\Delta} + \varepsilon_{\rm{AC}}, \quad
 \varepsilon_2 = \Delta, \quad
 \varepsilon_{3} = - \varepsilon_{\rm{AC}} .
\end{equation}
where
$\varepsilon_{\rm{AC}}=\frac{1}{2}\left(\sqrt{\Delta^2+g_1^2+g_2^2}-\Delta\right)$
is the ac-Stark shift. Note that we have labelled the dressed
states such that for $g_1 \gg g_2$ the state $|\varphi_i\rangle$
has the main weight on the electronic state $|i\rangle$, and that
$|\varphi_2\rangle$ represents the dark state which decouples from
the laser fields for $\eta=0$. It follows from the above equations
that the levels $\varepsilon_{2,n}$ and $\varepsilon_{1,n\!-\!1}$
cross each other (see Fig.~\ref{figure2}b) if the resonance
condition
\begin{equation}
\label{resonance}
 \omega =
 \varepsilon_{\rm{AC}}
 \approx \frac{g_1^2+g_2^2}{4\Delta}
\end{equation}
is satisfied. The approximation in Eq.~\eqref{resonance} refers to
the case $g_{1,2}\ll\Delta$ which is assumed throughout the paper.

For non-vanishing Lamb-Dicke parameter, $\eta>0$,  the electronic
and the translational degrees of freedom of the atom are coupled.
This coupling lifts the degeneracy of the levels
$\varepsilon_{2,n}$ and $\varepsilon_{1,n-1}$ at the crossing
point to yield an avoided crossing or Landau-Zener splitting
\cite{LANDAU,ZENER,STUECKELBERG} with a certain energy gap $\Delta
E_n$, as is illustrated in Fig.~\ref{figure2}b.

\begin{figure}[h]
\includegraphics[width=1\columnwidth]{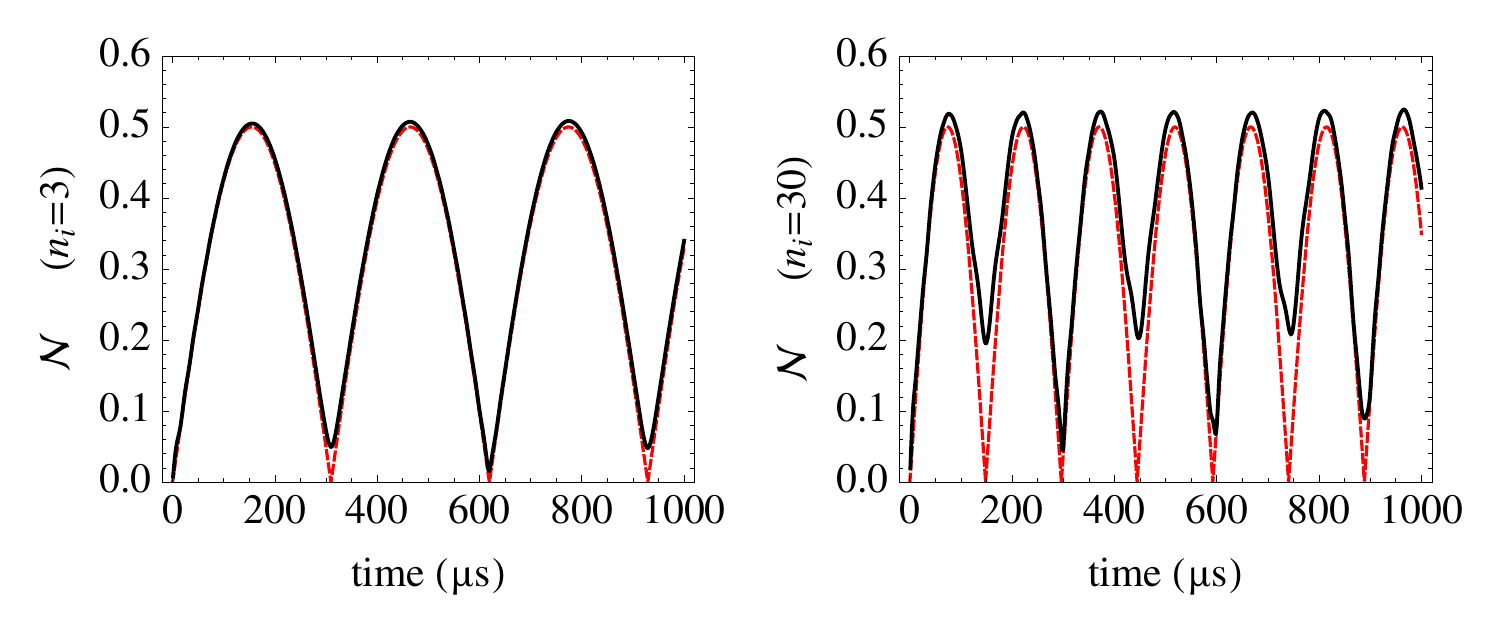}
\caption{Time dependence of the negativity obtained from the
Schr\"odinger equation \eqref{SCHROEDINGER} for initial
vibrational levels $n_{\rm{i}}=3$ and $30$. Parameters:
$\eta$=0.1, $\omega$=0.03, $\Delta$=15, $g_1$=1.34, $g_1/g_2$=10
(all frequencies are given in units of $2 \pi$\,MHz). The broken
lines represent the prediction according to
Eq.~\eqref{negativity-1}.} \label{figure3}
\end{figure}

Employing degenerate perturbation theory it can be shown that the
eigenstates at the center of the avoided crossing are given by the
even and odd linear combinations of the unperturbed (crossing)
states. More precisely, this means that with an appropriate choice
of the phase of the states $|\varphi_2,n\rangle$ and
$|\varphi_1,n-1\rangle$ the eigenstates of the full Hamiltonian at
resonance have the approximate form
\begin{equation} \label{LZ-pair}
 |\chi_{\pm}\rangle = \frac{1}{\sqrt{2}}\Big(
 |\varphi_2\rangle \otimes |n\rangle \pm |\varphi_1\rangle \otimes |n-1\rangle\Big)
\end{equation}
with the corresponding energy eigenvalues
\begin{equation}
 E_{\pm} = \Delta + n\omega \pm \frac{\Delta E_n}{2}.
\end{equation}
Considering as initial state
$|\psi(0)\rangle=|\varphi_2,n\rangle$, it follows that the
solution of the Schr\"odinger equation is given by
\begin{eqnarray} \label{state-1}
 |\psi(t)\rangle &=& \frac{1}{\sqrt{2}}\Big(
 e^{-iE_+t}|\chi_+\rangle + e^{-iE_-t}|\chi_-\rangle \Big) \nonumber \\
 &=& \frac{1}{2} \left( e^{-iE_+t} + e^{-iE_-t} \right)
 |\varphi_2\rangle\otimes|n\rangle \nonumber \\
 &+& \frac{1}{2} \left( e^{-iE_+t} - e^{-iE_-t} \right)
 |\varphi_1\rangle\otimes|n-1\rangle.
\end{eqnarray}
The Schmidt coefficients \cite{BENGTSSON} of this state take the
form
\begin{equation}
 \alpha_{1,2} = \frac{1}{2}\left|e^{-iE_+t}\pm e^{-iE_-t}\right|,
\end{equation}
which immediately leads to the negativity:
\begin{equation} \label{negativity-1}
 {\mathcal{N}}(t) = \alpha_1\alpha_2 = \frac{1}{2}|\sin(\Delta E_n \cdot t)| \, .
\end{equation}
The same formula is obtained if one considers the initial state
$|\psi(0)\rangle=|\varphi_1,n-1\rangle$. Thus, we see that the
entanglement dynamics is determined by a single spectral quantity,
namely the energy gap $\Delta E_n$, describing an oscillation of
the negativity with the period $T=\pi/\Delta E_n$. In
Fig.~\ref{figure3} we compare the negativity obtained from the
numerical solution of the full Schr\"odinger equation
(\ref{SCHROEDINGER}) with the formula \eqref{negativity-1} for
various initial states, which clearly demonstrates the accuracy of
the approximation based on a single Landau-Zener state pair
\eqref{LZ-pair}. The negativity predicted by
Eq.~\eqref{SCHROEDINGER} slightly exceeds the value $1/2$ as it
also accounts for the small contributions to entanglement by the
excited state levels $|\varphi_3,n\rangle$. We emphasize that the
argument which leads to the result \eqref{negativity-1} is quite
general and should be applicable to many different physical
scenarios: It only requires the existence of a Landau-Zener
splitting between tensor product states on some composite Hilbert
space. Moreover, depending only on the energy gap, formula
\eqref{negativity-1} can be applied even if the structure of the
factors of the tensor products is not known.

Next we discuss the influence of dissipation on the entanglement
dynamics which is described by the Lindblad dissipator of
Eq.~\eqref{LINDBLAD}. Figure~\ref{figure1} shows examples for the
negativity dynamics obtained from numerical solutions of the full
master equation \eqref{QMEQ}. We see that the negativity follows a
damped oscillation for small $\Gamma$, where the revivals of the
negativity come in pairs of nearly equal heights. At a certain
threshold value of $\Gamma$ the oscillations die out and a
monotonic decay of the negativity after the first bump is
observed.

The entanglement behavior under dissipation can be modelled
through the Schr\"odinger equation
\begin{equation} \label{SCHROEDINGER-EFF}
 \frac{d}{dt}|\psi(t)\rangle = -{\rm{i}}H_{\rm eff}^{\eta}|\psi(t)\rangle\, ,
\end{equation}
with the effective Hamiltonian
\begin{equation} \label{H-EFF}
 H_{\rm eff}^{\eta} = H^{\eta} - {\rm{i}}\frac{\Gamma}{2} |3\rangle\langle 3| \, .
\end{equation}
This is a non-Hermitian Hamiltonian derived from the quantum
master equation~\eqref{QMEQ}: It describes the coherent motion
generated by $H^{\eta}$ as well as the loss term (anti-commutator)
of the dissipator~\eqref{LINDBLAD} which leads to finite widths of
the dressed states of the atom-field system given by the imaginary
parts of the complex eigenvalues. Employing the effective
Schr\"odinger equation~\eqref{SCHROEDINGER-EFF} we thus neglect
the gain terms of the dissipator~\eqref{LINDBLAD}. These latter
terms describe real emission processes and induce, at least
partially, transitions into other Landau-Zener state pairs which
add incoherently and, thus, lead only to negligible contributions
to the negativity.

\begin{figure}[h]
\includegraphics[width=1\columnwidth]{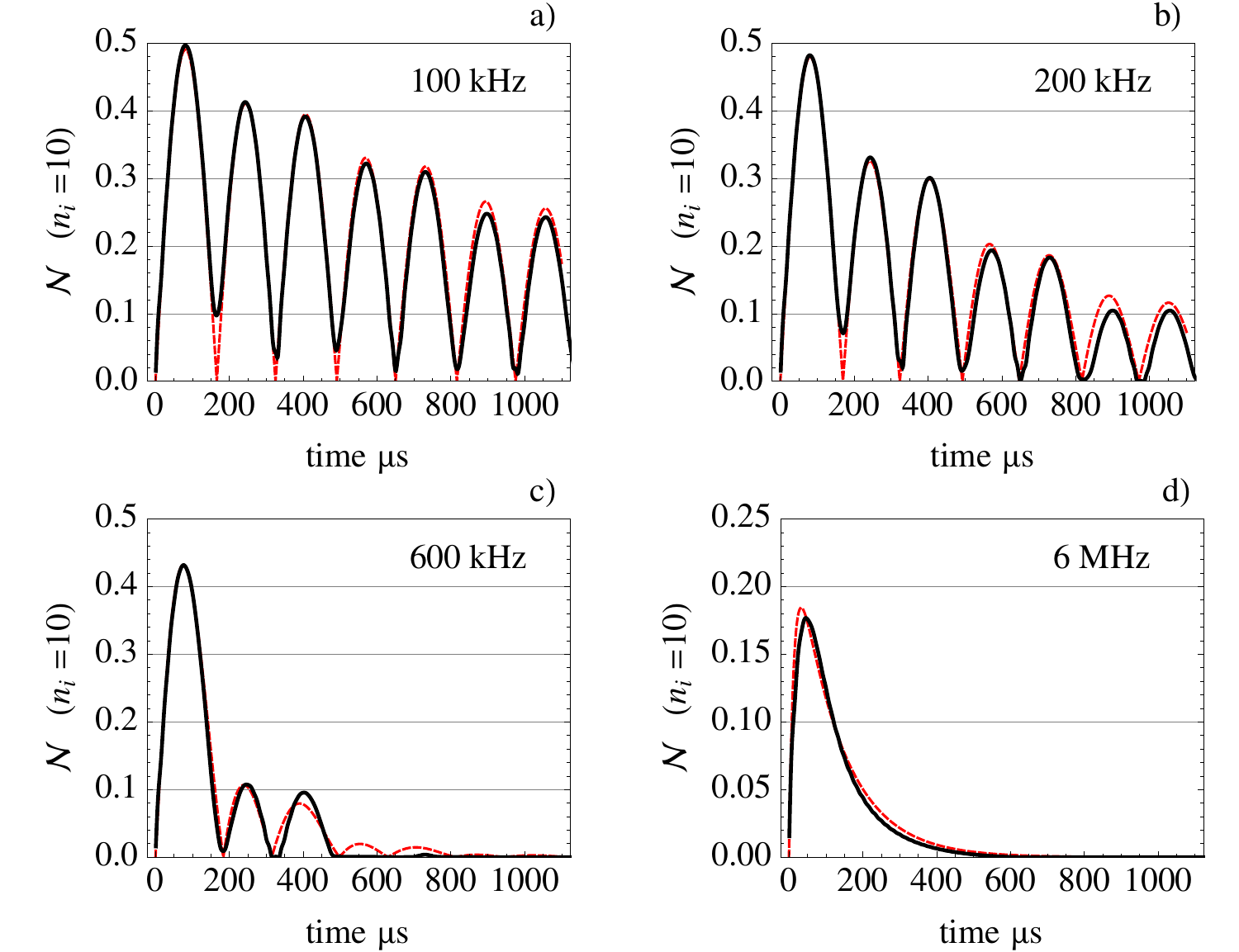}
\caption{The negativity obtained from the quantum master equation
(\ref{QMEQ}) for various $\Gamma$, as also shown in
Fig.~\ref{figure1}. Fits of Eq.~(\ref{negativity-2}) are shown by
the broken lines. Conditions as given in Fig.~\ref{figure3}.}
\label{figure4}
\end{figure}

In the framework of this approximation we can extend the
Landau-Zener state pair calculation to the effective Hamiltonian~\eqref{H-EFF}. Under the conditions $g_{1,2}\ll \Delta$ and
$\Gamma \ll \Delta$ the eigenvalues of $H_{\rm eff}^0$  are
\begin{eqnarray}
 \varepsilon_{1} &=& \Delta + \frac{g_1^2+g_2^2}{4\Delta} -{\rm{i}}\frac{\gamma_1}{2}, \\
 \varepsilon_2 &=& \Delta, \\
 \varepsilon_{3} &=& -\frac{g_1^2+g_2^2}{4\Delta}
 -\frac{\rm{i}}{2}(\Gamma-\gamma_1),
\end{eqnarray}
where
 $\gamma_1=\Gamma(g_1^2+g_2^2)/(4\Delta^2)$. We denote the
eigenstates corresponding to the eigenvalues $\varepsilon_{1,2}$
again by $|\varphi_{1,2}\rangle$, and the Landau-Zener pair by
$|\varphi_1,n-1\rangle$ and $|\varphi_2,n\rangle$. Using these
states as basis states, the effective Hamiltonian \eqref{H-EFF}
reduces within the two-state resonance approximation to the matrix
\begin{equation} \label{H-EFF-APPROX}
 H_{\rm eff}^{\eta} = \left(
 \begin{array}{cc}
 \Delta + n\omega -{\rm{i}}\gamma_1/2 & \Delta E_n/2 \\
 \Delta E_n/2 & \Delta + n\omega
 \end{array} \right).
\end{equation}
This Hamiltonian may be viewed as describing damped Rabi-type
oscillations between the Landau-Zener pair states with Rabi frequency
$\Delta E_n$, where only the state $|\varphi_1,n-1\rangle$ is
damped, since the width of the dark state $|\varphi_2,n\rangle$ is
practically zero. With the effective Hamiltonian
\eqref{H-EFF-APPROX} it is now easy to determine the solution of
the Schr\"odinger equation \eqref{SCHROEDINGER-EFF} corresponding
to the initial state $|\psi(0)\rangle=|\varphi_2,n\rangle$, and to
derive 
for the negativity,
\begin{equation} \label{negativity-2}
 {\mathcal{N}}(t) = \frac{\Delta E_n}{2} \left| \frac{\sin\nu t}{\nu}
 + \frac{\gamma_1}{2\nu^2} (1-\cos \nu t) \right| e^{-\gamma_1 t/2},
\end{equation}
where $\nu=\sqrt{(\Delta E_n)^2-\gamma_1^2/4}$. As is illustrated
in Fig.~\ref{figure4} the simple formula \eqref{negativity-2}
provides an excellent approximation of the entanglement dynamics
for $\gamma_1/2 < \Delta E_n$, and even for intermediate values of
the spontaneous emission rate the qualitative behavior of the
negativity is reproduced.
It is remarkable that
Eq.~\eqref{negativity-2} correctly describes the transition from
the underdamped to the overdamped motion of the negativity at
$\gamma_1/2=\Delta E_n$, compare Figs.~\ref{figure4}c and
\ref{figure4}d. Using the approximations $\Delta E_n \approx
\eta\sqrt{n}g_1g_2/\Delta$ and
 $\gamma_1 \!\approx \!\Gamma
g_1^2/(4\Delta^2)$ this transition is predicted to occur at the
value 
$\Gamma = 8\eta\sqrt{n}\Delta g_2/g_1$, which nicely fits to
the results of our numerics.

The long-time behavior of the negativity for larger values of the
spontaneous emission rate $\Gamma$ exhibits an interesting and
complex structure, see Fig.~\ref{figure1}.  The exponential decay
of the initial peak of the negativity is followed by a precipitous
drop until the rate of decay slows down and a smooth transition to
an algebraic decay of the form ${\mathcal{N}}(t) \propto t^{-3/2}$
emerges, see the broken line in Fig.~\ref{figure1}. While a
detailed discussion is beyond the scope of this note, we mention
that the Landau-Zener picture can also offer an explanation here.
A power law governs the negativity dynamics until a rebirth of
entanglement towards the final value
${\mathcal{N}}(\infty)={\mathcal{N}}(\rho_{\rm stat})$ occurs. The
final value of the negativity represents the entanglement in the
unique non-equilibrium stationary state $\rho_{\rm stat}$ of the
master equation \eqref{QMEQ}. During the cooling the atom is
driven into a mixed stationary state which has a dominant
contribution from the state $|\varphi_2,0\rangle$, the product of
the dark state and the vibrational ground state of the trap.
However, $\rho_{\rm stat}$ contains a small amount of entanglement
which is controlled by the small steady-state heating caused by
off-diagonal transitions from the dark state. Also here the
LZ-model offers a quantitative explanation. The interaction of the
lowest dark state, $\langle \varphi_{2},0| H_{\rm{int}}^{\eta}
|\varphi_{1},1\rangle\! = \!\eta \, g_1g_2/(4\Delta)$, leads to a
steady state negativity, ${\mathcal{N}}(\infty)$=$\eta \, g_2/g_1$
for $\Gamma\! \to \!0$, in agreement with the predictions from
Eq.~(\ref{QMEQ}) for the examples  in Fig.~\ref{figure1}.

The Landau-Zener mechanism suggest various methods for the control
of entanglement between electronic and translational degrees of
freedom of trapped atoms by means of suitable laser fields. One
possibility is  the switching of fields at appropriate points of
time, e.g., at the instant of maximal entanglement. Since the
manifold spanned by the Landau-Zener state pair has only a
negligible contribution from the excited electronic state, the
negativity practically freezes upon rapidly switching-off the
lasers.  This method can thus create permanent entanglement of the
trapped atom. A further possibility is to use suitable laser
pulses in order to generate entanglement through the control of
adiabatic/diabatic Landau-Zener transitions. A detailed discussion
of these points will be given in a forthcoming paper.

\acknowledgments This work was supported by the Deutsche
Forschungsgemeinschaft (Grant HE-2525/8).

\end{document}